\documentclass[12pt]{article}
\usepackage{color}
\usepackage{amssymb,amsmath,comment}
\usepackage{graphicx}
\usepackage{braket}
\usepackage{epsfig}
\usepackage{here}
\usepackage{bm}
\graphicspath{{./Figures/}}
\newcommand{\ov}{\overline}
\newcommand{\ba}{\begin{align}}
\newcommand{\ea}{\end{align}}

\def\nn{\nonumber}

\def\bea{\begin{eqnarray}}
\def\eea{\end{eqnarray}}
 \makeatletter
\def\alt{\mathrel{\mathpalette\gl@align<}}
\def\agt{\mathrel{\mathpalette\gl@align>}}
\def\gl@align#1#2{\lower.6ex\vbox{\baselineskip\z@skip\lineskip\z@
\ialign{$\m@th#1\hfil##\hfil$\crcr#2\crcr\sim\crcr}}} \makeatother
\renewcommand{\thefootnote}{\fnsymbol{footnote}}
\begin{document}
\begin{flushright}
\end{flushright}
\vspace*{1.0cm}

\begin{center}
\baselineskip 20pt 
{\Large\bf 
Are Low-energy Data already Hinting at Five Dimensions?
}
\vspace{1cm}

{\large 
Naoyuki Haba$^a$, Nobuchika Okada$^b$ and Toshifumi Yamada$^a$
} \vspace{.5cm}

{\baselineskip 20pt \it
$^a$ Institute of Science and Engineering, Shimane University, Matsue 690-8504, Japan

$^b$ Department of Physics and Astronomy, University of Alabama, Tuscaloosa, Alabama 35487, USA
}

\vspace{.5cm}

\vspace{1.5cm} {\bf Abstract} \end{center}

Low-energy data, combined with renormalization group (RG) equations, can predict new physics at far higher energy scales.
In this paper, we consider the possibility that the measured Higgs boson mass and top quark mass hint at 
 a five-dimensional gauge-Higgs unification (5D GHU) model at a scale above TeV.
We note that the vanishing of the Higgs quartic coupling and the proximity of the top quark Yukawa coupling and weak gauge coupling
 at high scales, inferred from the experimental data, are in harmony with 5D GHU,
 because in 5D GHU models the Higgs quartic coupling is forbidden by the 5D gauge symmetry and 
 the Yukawa couplings and the weak gauge coupling originate from a common 5D gauge coupling.
Based on the above insight, we propose a 5D GHU model 
 where the Standard Model fermions are embedded in 5D fermions in a way to tightly relate the top Yukawa coupling
 with the weak gauge coupling.
Also, the model predicts the presence of vector-like fermions (other than the Kaluza-Klein modes),
 which can affect the RG evolutions of the 4D theory and reconcile the scale of vanishing Higgs quartic coupling and that of equality of the top Yukawa and weak gauge couplings, thereby achieving a successful matching of the 4D theory with 5D GHU.
We predict the vector-like fermion mass and the compactification scale of 5D GHU from the conditions for the successful matching.

\thispagestyle{empty}

\newpage
\renewcommand{\thefootnote}{\arabic{footnote}}
\setcounter{footnote}{0}
\baselineskip 18pt

\section{Introduction}

In the Standard Model (SM), the Higgs quartic coupling
 exhibits an interesting property that it vanishes at an energy scale below the Planck scale
 along renormalization group (RG) evolutions.
This property is usually interpreted as implying metastability of our vacuum~\cite{Buttazzo:2013uya}
 as the Higgs quartic coupling turns negative above that scale.
Another interpretation~\cite{Haba:2005kc,Haba:2008dr,Carson:2015ova,Haba:2016xji} is that 
 the SM is matched, around the scale of vanishing Higgs quartic coupling, 
 with a model of gauge-Higgs unification~\cite{Manton:1979kb,Fairlie:1979at,Fairlie:1979zy,Hosotani:1983xw,Hosotani:1983vn,Hosotani:1988bm}
 \footnote{
For early work on extradimensional models, see, e.g., Refs.~\cite{Antoniadis:1990ew,Antoniadis:1993jp}.
 } in five dimensions (5D GHU), where
 the SM Higgs field is identified with the fifth dimensional component of a gauge field whose quartic coupling is forbidden by the 5D gauge symmetry.

Although less noted, the top quark Yukawa coupling and the weak gauge coupling also show an interesting feature
 that they become equal at the energy scale of $O(10^8)$~GeV.
This suggests that the two couplings may have a common origin in some new physics above that scale.
Indeed, in 5D GHU, the SM Yukawa interactions are part of an extension of the weak interaction
 and the Yukawa couplings are tied with the weak gauge coupling.

Motivated by the above insights, we propose a model in which 
 the scale where the Higgs quartic coupling vanishes is reconciled with the scale where the top quark Yukawa coupling equals the weak gauge coupling,
 and this common scale is interpreted as the scale where 5D GHU emerges.

Our model is characterized by the direct embedding of the SM fermions into 5D bulk fermions.
Here, we introduce 5D fermions in {\bf 3} representation of $SU(3)_W$ gauge group that extends the $SU(2)_W$ weak gauge group.
After orbifold compactification, one 5D fermion yields one isospin-doublet and one isospin-singlet 4D fermions of opposite chiralities 
 as the massless modes,
 which are directly identified with an isospin-doublet and an isospin-singlet SM quarks or leptons.
Consequently, all the SM Yukawa couplings are identical with the weak gauge coupling at this stage.
Small Yukawa couplings other than the top quark Yukawa coupling
 are reproduced by introducing 4D vector-like fermions at an orbifold fixed point that mix with the massless modes of 5D fermions
\footnote{
In the next section, we will show that it is the square sum of the up-type and down-type quark Yukawa matrices that is related to the weak gauge coupling.
Hence, the bottom quark Yukawa coupling can be small without help of the additional 4D vector-like fermions.
}.
The above structure materializes the idea that the Yukawa couplings and the weak gauge coupling are basically the same entity
 and this fact is encoded in the value of the top quark Yukawa coupling.
Thus, this structure is aligned with our motivation to interpret the scale of equality of the top Yukawa coupling and weak gauge coupling, as the scale of 5D GHU.

Another characteristic of our model is the presence of vector-like fermions (other than Kaluza-Klein modes)
 that is not arbitrary but is required by the model structure.
They are important not only theoretically, but also phenomenologically;
 since their mass can be smaller than the compactification scale, 
 they can change the RG evolutions of the 4D theory and achieve its matching with the 5D GHU theory.
In fact, they can amend a discrepancy in the SM between
 the scale of equality of the top Yukawa and weak gauge couplings and
 the scale of vanishing Higgs quartic coupling,
 thereby making the matching of the 4D and 5D theories successful.

It should be noted that our model is not intended to solve the hierarchy problem of the Higgs mass,
 and the compactification scale can be much larger than TeV scale.

Previously, the unification of the weak gauge coupling and top quark Yukawa coupling in a high-scale 5D GHU model
 has been discussed in Ref.~\cite{Abdalgabar:2017cjw}, but only in a toy model.
We in this paper investigate the unification of the two couplings in a complete, realistic model,
 and further relate the unification with the high-scale vanishing of the Higgs quartic coupling.

This paper is organized as follows.
In Section~\ref{model}, we present our model.
In Section~\ref{analysis}, we determine the mass of the vector-like fermions and the compactification scale of 5D GHU,
 by solving the RG equations of the 4D theory made of the SM content and the vector-like fermions,
 and matching it with the 5D GHU theory.
Section~\ref{summary} summarizes the paper.
\\

\section{Model}
\label{model}

We consider $SU(3)_C\times SU(3)_W\times U(1)_V$ gauge theory in a flat 5D spacetime 
 whose fifth dimension is compactified on $S^1/Z_2$ orbifold.
We write the fifth coordinate as $y$, and
the $S^1$ is obtained by the identification of $y$ with $y+2\pi R$.
The $S^1/Z_2$ orbifold is then obtained by identifying $y$ with $-y$.
$SU(3)_C$ is the SM QCD gauge group, and $SU(3)_W\times U(1)_V$ incorporates the electroweak gauge groups.
We denote the $SU(3)_W$ gauge field by $(W_\mu,W_5)$ and the $U(1)_V$ gauge field by $(V_\mu,V_5)$, where $\mu=0,1,2,3$.
The fields transform under the $Z_2$ as
\begin{align}
W_\mu(y)&=PW_\mu(-y)P^\dagger, \ \ \ \ \ W_5(y)=-PW_5(-y)P^\dagger,
\nn\\
V_\mu(y)&=V_\mu(-y), \ \ \ \ \ V_5(y)=-V_5(-y),
\nn\\
& \ \ \ \ \  \ \ P=\begin{pmatrix} 
      -1& 0 & 0 \\
      0 & -1 & 0 \\
            0 & 0 & 1 \\
   \end{pmatrix},
   \label{z2}
\end{align}
 and accordingly $SU(3)_W\times U(1)_V$ is broken into $SU(2)_W\times U(1)_W\times U(1)_V$ at $y=0,\pi R$.
Gauge group $U(1)_W\times U(1)_V$ is further broken into SM hypercharge $U(1)_Y$ by a brane-localized scalar field
 in $({\bf 1},{\bf 1},-\frac{1}{2\sqrt{3}},\frac{1}{2\sqrt{3}})$ representation of $SU(3)_C\times SU(2)_W\times U(1)_W\times U(1)_V$,
 denoted by $\phi$, that develops a vacuum expectation value (VEV).
Accordingly, the hypercharge $Q_Y$ is given in terms of the $U(1)_W$ charge $Q_W$ and the $U(1)_V$ charge $Q_V$ by
\bea
Q_Y=\frac{1}{\sqrt{3}}(Q_W+Q_V).
\label{u1}
\eea
The four-dimensional components of the $SU(2)_W\times U(1)_Y$ gauge fields have massless modes, which are identified with the SM electroweak gauge fields.
The fifth dimensional component of the $SU(3)_W/(SU(2)_W\times U(1)_W)$ gauge field has massless KK mode, which is identified with the SM Higgs field,
 $H$, as
\bea
W_5|_{\rm massless \ KK} \ = \    \frac{1}{\sqrt{2}}\begin{pmatrix} 
      O & H \\
      H^\dagger & 0 \\
   \end{pmatrix}.
\label{higgs}
\eea

We introduce three copies of four bulk 5D fermions in 
 $({\bf \bar{3}},{\bf 3},-\frac{1}{\sqrt{3}})$, $({\bf 3},{\bf 3},0)$, $({\bf 1},{\bf 3},-\frac{2}{\sqrt{3}})$, $({\bf 1},{\bf 3},\frac{1}{\sqrt{3}})$ 
 representations of $SU(3)_C\times SU(3)_W\times U(1)_V$,
 denoted by $\Psi_{u^c}^i$, $\Psi_{d}^i$, $\Psi_{e}^i$, $\Psi_{\nu^c}^i$ with $i=1,2,3$.
The three copies correspond to the three generations of the SM.
The 5D fermions transform under the $Z_2$ as
\footnote{
Since $\Psi_{u^c}^i$, $\Psi_{d}^i$, $\Psi_{e}^i$, $\Psi_{\nu^c}^i$ are in {\bf 3} representation of $SU(3)_W$,
 the same $P$ as Eq.~(\ref{z2}) enters Eqs.~(\ref{uc})-(\ref{nuc}).
}
\bea
\Psi_{u^c}^i(-y) &=& -\gamma_5P\,\Psi_{u^c}^i(y),
\label{uc}\\
\Psi_{d}^i(-y) &=& \gamma_5P\,\Psi_{d}^i(y),
\\
\Psi_{e}^i(-y) &=& \gamma_5P\,\Psi_{e}^i(y),
\\
\Psi_{\nu^c}^i(-y) &=& -\gamma_5P\,\Psi_{\nu^c}^i(y).
\label{nuc}
\eea
The 5D fermions are summarized in Table~\ref{bulkfermions}.
\begin{table}[H]
\begin{center}
  \caption{The gauge and $Z_2$ charges of the 5D fermions. $i$ is the flavor index with $i=1,2,3$.}
  \begin{tabular}{|c|c|c|c|c|} \hline
                             & $SU(3)_C$ & $SU(3)_W$ & $U(1)_V$ & $Z_2$ transformation                        \\ \hline 
                             & & & & \\
    $\Psi_{u^c}^i$    & ${\bf \bar{3}}$ & ${\bf 3}$ & $-\frac{1}{\sqrt{3}}$ & $\Psi_{u^c}^i(-y)=-\gamma_5P\,\Psi_{u^c}^i(y)$                     \\
    $\Psi_{d}^i$       & ${\bf 3}$ & ${\bf 3}$ & $0$ & $\Psi_{d}^i(-y)=\gamma_5P\,\Psi_{d}^i(y)$                     \\ 
    $\Psi_{e}^i$       & ${\bf 1}$ & ${\bf 3}$ & $-\frac{2}{\sqrt{3}}$ & $\Psi_{e}^i(-y)=\gamma_5P\,\Psi_{e}^i(y)$                     \\
    $\Psi_{\nu^c}^i$  & ${\bf 1}$ & ${\bf 3}$ & $\frac{1}{\sqrt{3}}$ & $\Psi_{\nu^c}^i(-y)=-\gamma_5P\,\Psi_{\nu^c}^i(y)$                     \\ \hline
  \end{tabular}
  \label{bulkfermions}
  \end{center}
\end{table}
\noindent
From the $Z_2$ charge assignments, we find that the following chiral fields with $SU(3)_C\times SU(2)_W\times U(1)_W\times U(1)_V$ charges
 possess massless KK modes:
\bea
{\rm Right\mathchar`-handed \ \ } \left({\bf \bar{3}},{\bf 2},\frac{1}{2\sqrt{3}},-\frac{1}{\sqrt{3}}\right) {\rm \ \ component \ of \ }\Psi_{u^c}^i
\label{ur0}\\
{\rm Left\mathchar`-handed \ \ } \left({\bf \bar{3}},{\bf 1},-\frac{1}{\sqrt{3}},-\frac{1}{\sqrt{3}}\right) {\rm \ \ component \ of \ }\Psi_{u^c}^i
\\
{\rm Left\mathchar`-handed \ \ } \left({\bf 3},{\bf 2},\frac{1}{2\sqrt{3}},0\right) {\rm \ \ component \ of \ }\Psi_{d}^i
\label{dl0}\\
{\rm Right\mathchar`-handed \ \ } \left({\bf 3},{\bf 1},-\frac{1}{\sqrt{3}},0\right) {\rm \ \ component \ of \ }\Psi_{d}^i
\\
{\rm Left\mathchar`-handed \ \ } \left({\bf 1},{\bf 2},\frac{1}{2\sqrt{3}},-\frac{2}{\sqrt{3}}\right) {\rm \ \ component \ of \ }\Psi_{e}^i
\label{el0}\\
{\rm Right\mathchar`-handed \ \ } \left({\bf 1},{\bf 1},-\frac{1}{\sqrt{3}},-\frac{2}{\sqrt{3}}\right) {\rm \ \ component \ of \ }\Psi_{e}^i
\\
{\rm Right\mathchar`-handed \ \ } \left({\bf 1},{\bf 2},\frac{1}{2\sqrt{3}},\frac{1}{\sqrt{3}}\right) {\rm \ \ component \ of \ }\Psi_{\nu^c}^i
\label{nur0}\\
{\rm Left\mathchar`-handed \ \ } \left({\bf 1},{\bf 1},-\frac{1}{\sqrt{3}},\frac{1}{\sqrt{3}}\right) {\rm \ \ component \ of \ }\Psi_{\nu^c}^i
\label{nul0}
\eea
At this stage, there are chiral anomalies with respect to the $U(1)_V$ gauge group.
These anomalies should be cancelled, and to achieve the cancellation, we introduce
 4D chiral fermions of Table~\ref{localizedfermions} at an orbifold fixed point $y=0$:
\begin{table}[H]
\begin{center}
  \caption{The gauge charges and chirality of the 4D fermions localized at $y=0$. $i$ is the flavor index with $i=1,2,3$.}
  \begin{tabular}{|c|c|c|c|c|c|} \hline
                             & $SU(3)_C$ & $SU(2)_W$ & $U(1)_W$ & $U(1)_V$ & chirality                        \\ \hline
                             & & & & & \\
    $\psi_{qR}^i$      & ${\bf 3}$ & ${\bf 2}$ & 0 & $\frac{1}{2\sqrt{3}}$ & Right-handed                     \\
    $\psi_{\ell R}^i$  & ${\bf 1}$ & ${\bf 2}$ & 0 & $-\frac{3}{2\sqrt{3}}$ & Right-handed                     \\ \hline
  \end{tabular}
  \label{localizedfermions}
  \end{center}
\end{table}
\noindent
The presence of 4D fermions $\psi_{qR}^i$, $\psi_{\ell R}^i$ is probably a consequence of more fundamental physics that underlies the orbifolding.
In the present paper, however, we adhere to the orbifold picture and introduce the 4D fermions by hand.

The most generic 5D action is
\begin{align}
S=\int{\rm d}^4x\int^{\pi R}_{-\pi R}{\rm d}y \ &-\frac{1}{2}{\rm tr}\left[W_{MN} W^{MN}\right]-\frac{1}{4}V_{MN}V^{MN}
\nn\\
&+i\overline{\Psi}_{u^c}^i\,\Gamma^MD_M\Psi_{u^c}^i +i\overline{\Psi}_{d}^i\,\Gamma^MD_M\Psi_{d}^i 
  +i\overline{\Psi}_{e}^i\,\Gamma^MD_M\Psi_{e}^i +i\overline{\Psi}_{\nu^c}^i\,\Gamma^MD_M\Psi_{\nu^c}^i
\nn\\
&+\delta(y) \left[ i\psi_{qR}^{i\,\dagger}\,\bar{\sigma}^\mu D_\mu \,\psi_{qR}^i +  i\psi_{\ell R}^{i\,\dagger}\,\bar{\sigma}^\mu D_\mu \,\psi_{\ell R}^i \right.
\nn\\
&\ \ \  -\psi_{qR}^{i\,T}\left\{A_{ij}\ \phi\ i\sigma_2\,\Psi_{u^c}^j|_R + B_{ij}\ \phi^\dagger\  \Psi_{d}^j|_L^*\right\}
-\psi_{\ell R}^{i\,T}\left\{C_{ij}\  \phi^\dagger \ \Psi_{e}^j|_L^* + E_{ij}\ \phi\  i\sigma_2\,\Psi_{\nu^c}^j|_R\right\}
\nn\\
&\left. \ \ \  + {\rm H.c.} \ \right]
\label{action}
\end{align}
 where spacetime indices $M,N$ run as $M,N=0,1,2,3,5$,
 and $D_M,D_\mu$ denote gauge-covariant derivatives.
$\Psi_{u^c}^j|_R$, $\Psi_{d}^j|_L$, $\Psi_{e}^j|_L$, $\Psi_{\nu^c}^j|_R$ denote the 4D-decomposed components of
 the 5D bulk fermions
 with the indicated chirality that are in $SU(2)_W$ doublet representation, listed in Eqs.~(\ref{ur0}),(\ref{dl0}),(\ref{el0}),(\ref{nur0}).
$A_{ij},B_{ij},C_{ij},E_{ij}$ denote their coupling constants with 4D localized fermions 
 $\psi_{qR}^i,\psi_{\ell R}^i$ and the $U(1)_W\times U(1)_V$-breaking scalar $\phi$.
After $\phi$ develops a VEV, the fourth line of Eq.~(\ref{action}) yields a quark mass matrix,
 which can be recast by a flavor rotation of $\psi_{qR}^i$ into 
\bea
      \psi_{qR}^{T}
   \begin{pmatrix} 
     M_{3\times3} & O_{3\times3}
   \end{pmatrix}
      \begin{pmatrix} 
         U_1 & U_2 \\
         U_3 & U_4 \\
      \end{pmatrix}
   \begin{pmatrix} 
      i\sigma_2\,\Psi_{u^c}|_R \\
      \Psi_{d}|_L^* \\
   \end{pmatrix}
\label{massmatrix}
\eea
 where $M_{3\times3}$ is a $3\times3$ diagonal matrix,
 $O_{3\times3}$ denotes the $3\times3$ null matrix,
 $\begin{pmatrix} 
         U_1 & U_2 \\
         U_3 & U_4 \\
      \end{pmatrix}$
  is a unitary matrix, and $U_1,U_2,U_3,U_4$ are its $3\times3$ submatrices.
We see that the fields below are massless chiral modes,
\bea
(U_3)_{kj} \, i\sigma_2\,\Psi_{u^c}^j|_R + (U_4)_{kj} \, \Psi_{d}^j|_L^* \ \equiv \ Q_L^{k\,*}.
\label{chiraldoublet}
\eea
$Q_L^{k}$ are identified with the SM isospin-doublet quarks. 
 (Remember that $\Psi_{u^c}^j|_R$, $\Psi_{d}^j|_L^*$ have hypercharge $Q_Y=1/6$ after the breaking of $U(1)_W\times U(1)_V$ into $U(1)_Y$.)
There also are vector-like modes with mass $M_{3\times3}$ that comprise
\bea
&&\psi_{qR}^k \ \ \ {\rm and } \ \ \ \ (U_1)_{kj} \, i\sigma_2\,\Psi_{u^c}^j|_R + (U_2)_{kj} \, \Psi_{d}^j|_L^* \equiv X_{qL}^{k\,*}.
\label{vecquarks}
\eea
They represent new, vector-like quarks, which have the same charges as the SM isospin-doublet quarks.
Other fields $\Psi_{u^c}^j|_L$, $\Psi_{d}^j|_R$ are identified with the SM up-type and down-type isospin-singlet quarks.

The presence of the vector-like quarks Eq.~(\ref{vecquarks}) is an important prediction of the model.
As for theoretical aspects, their presence is not arbitrary but is required by the model structure.
Accordingly, their number and gauge charges are uniquely fixed.
As for phenomenological aspects, since their mass $M_{3\times 3}$ can be smaller than the compactification scale,
 they can modify the RG evolutions of SM parameters.
This modification can fix the discrepancy between the scale of vanishing Higgs quartic coupling and that of equality of the top Yukawa and weak gauge couplings, achieving the matching of the 4D theory with the 5D GHU theory.

The quark Yukawa interactions are derived from the 5D $SU(3)_W$ gauge interaction as
\bea
S \ \supset \ 2\pi R\int{\rm d}^4x \ i\frac{g_W}{\sqrt{2}} \, H \left\{ \Psi_{u^c}^{i}|_L^\dagger \, i\sigma_2 \ (U_3^\dagger)_{ik}\ Q_L^{k\,*}
+ \Psi_{d}^{i}|_R^T \ (U_4^\dagger)_{ik}\ Q_L^{k\,*}\right\}+{\rm H.c.},
\eea
 where $g_W$ denotes the $SU(3)_W$ gauge coupling.
The up-type and down-type quark Yukawa matrices, $Y_u,Y_d$, are extracted as 
$Y_u = i\frac{g_W}{\sqrt{2}}U_3^*$, $Y_d = -i\frac{g_W}{\sqrt{2}}U_4^*$.
Unfortunately, these Yukawa matrices are unrealistic because the unitarity relation $U_3^\dagger U_3+U_4^\dagger U_4=$diag$(1,1,1)$
 gives $Y_u^\dagger Y_u+Y_d^\dagger Y_d = \frac{g_W^2}{2}$diag$(1,1,1)$, which contradicts the smallness of the 1st and 2nd generation Yukawa couplings.
To reproduce their smallness, we introduce two generations of 4D vector-like fermions of Table~\ref{1st2nd} at $y=0$.
\begin{table}[H]
\begin{center}
  \caption{4D vector-like fermions localized at $y=0$.
  Index $a$ runs as $a=1,2$.}
  \begin{tabular}{|c|c|c|c|c|} \hline
                             & $SU(3)_C$ & $SU(2)_W$ & $U(1)_W$ & $U(1)_V$                         \\ \hline
                             & & & & \\
    $\chi_{u}^a$      & ${\bf 3}$ & ${\bf 1}$ & $\frac{1}{\sqrt{3}}$ & $\frac{1}{\sqrt{3}}$                      \\
    $\chi_{d}^a$     & ${\bf 3}$ & ${\bf 1}$ & $-\frac{1}{\sqrt{3}}$ & 0                      \\ \hline
  \end{tabular}
  \label{1st2nd}
  \end{center}
\end{table}
\noindent
4D vector-like fermions $\chi_{u}^a$, $\chi_{d}^a$ mix with the massless KK modes of 5D fermions through the term
\begin{align}
-\Delta S=\int{\rm d}^4x\int^{\pi R}_{-\pi R}{\rm d}y \ 
\delta(y) \left[ M_{u}^{aj} \ \chi_{u}^{a}|^T_L\, i\sigma_2 \,\Psi_{u^c}^j|_L   + M_{d}^{aj} \ \chi_{d}^{a}|_L^\dagger \,\Psi_{d}^j|_R + {\rm H.c.}
+ \mu_u^a\,\bar{\chi}_u^a\chi_u^a + \mu_d^a\,\bar{\chi}_d^a\chi_d^a
\right],
\label{extraaction}
\end{align}
 which reduces the Yukawa couplings of two generations by the ratio $\mu_u/M_u$ or $\mu_d/M_d$.

Even with the above mechanism to reduce the 1st and 2nd generation Yukawa couplings,
 there is a tight connection between the 3rd generation Yukawa couplings and the $SU(3)_W$ gauge coupling.
Since the relation $Y_u^\dagger Y_u+Y_d^\dagger Y_d = \frac{g_W^2}{2}$diag$(1,1,1)$ is still valid for the 3rd generation quarks,
 by neglecting the mixing between the 3rd generation and other generation quarks, we get
\bea
y_t^2+y_b^2 \ \simeq \ \frac{1}{2}g_W^2,
\label{y-w}
\eea
 where $y_t,y_b$ denote the top and bottom quark Yukawa couplings.
As $y_t$ is dominantly large, and $g_W$ matches with the weak gauge coupling, Eq.~(\ref{y-w}) materializes the idea that 
 equality of the top quark Yukawa coupling and the weak gauge coupling hints at 5D GHU.
\\

As for the lepton sector, the charged lepton Yukawa matrix and the neutrino Dirac Yukawa matrix are derived in the same fashion.
The fourth line of Eq.~(\ref{action}) gives a lepton mass matrix, which can be recast into the form
\bea
      \psi_{\ell R}^{T}
   \begin{pmatrix} 
     M'_{3\times3} & O_{3\times3}
   \end{pmatrix}
      \begin{pmatrix} 
         V_1 & V_2 \\
         V_3 & V_4 \\
      \end{pmatrix}
   \begin{pmatrix} 
      \Psi_{e}|_L^* \\
      i\sigma_2\,\Psi_{\nu^c}|_R \\
   \end{pmatrix}
\label{massmatrix2}
\eea
 where $M'_{3\times3}$ is a $3\times3$ diagonal matrix,
 $\begin{pmatrix} 
         V_1 & V_2 \\
         V_3 & V_4 \\
      \end{pmatrix}$
  is a unitary matrix, and $V_1,V_2,V_3,V_4$ are its $3\times3$ submatrices.  
The massless chiral modes below,
\bea
(V_3)_{kj} \,\Psi_{e}^j|_L^*  + (V_4)_{kj} \, i\sigma_2\,\Psi_{\nu^c}^j|_R \ \equiv \ L_L^{k\,*},
\eea 
 are identified with the SM isospin-doublet leptons.
The vector-like modes with mass $M'_{3\times3}$ that comprise
\bea
&&\psi_{\ell R}^k, \nn\\
&&(V_1)_{kj} \, (\Psi_{e}^j|_L)^* + (V_2)_{kj} \, i\sigma_2\,\Psi_{\nu^c}^j|_R \equiv X_{\ell L}^{k\,*}
\label{vecleptons}
\eea
 represent new, vector-like leptons, which have the same charges as the SM isospin-doublet leptons.
Other fields $\Psi_{e}^j|_R$, $\Psi_{\nu^c}^j|_L$ are identified with the SM charged leptons and isospin-singlet neutrinos, respectively.
The realistic charged lepton Yukawa couplings and tiny active neutrino masses
 are obtained by introducing vector-like fermions analogous to those of Table~\ref{1st2nd}.

As with the vector-like quarks, the presence of the vector-like leptons Eq.~(\ref{vecleptons}) is an important prediction of the model.
\\

\section{Estimation of Vector-like Fermion Mass and Compactification Scale}
\label{analysis}

The presence of vector-like quarks Eq.~(\ref{vecquarks}) and vector-like leptons Eq.~(\ref{vecleptons})
 can modify the RG evolutions of SM parameters and amend the discrepancy 
 between the scale of vanishing Higgs quartic coupling and that of equality of the top Yukawa and weak gauge couplings,
 enabling us to match the 4D theory with the 5D GHU theory.
Noting this fact, we estimate the mass of the vector-like fermions with which the matching conditions of the 4D theory with the 5D GHU theory are satisfied.
Additionally, we obtain the compactification scale from the matching scale.

The matching relates the radiatively-generated potential for the fifth dimensional component of $SU(3)_W$ gauge boson in 5D GHU
 with the Higgs potential in the 4D theory.
It also relates the $SU(3)_W$ gauge coupling in the former with the weak gauge coupling and quark Yukawa couplings in the latter.
For the Higgs potential, we adopt the general result~\cite{Haba:2005kc} that the scale of vanishing Higgs quartic coupling $\lambda$ at one loop 
 equals the compactification scale of 5D GHU models as
\bea
\lambda\left(\frac{1}{2\pi R}\right) \ = \ 0.
\label{comp-quartic}
\eea
Note that Eq.~(\ref{comp-quartic}) has been derived by considering one-loop threshold corrections from all the Kaluza-Klein modes of the $SU(3)_W$ gauge field and bulk fermions
 in a general 5D GHU model.
Hence, we can use Eq.~(\ref{comp-quartic}) without further including threshold corrections from Kaluza-Klein modes.
Note also that the 4D fermions localized at $y=0$ do not alter Eq.~(\ref{comp-quartic}) at one-loop level
 because these fermions do not couple to the Higgs field at tree level.
For the weak gauge coupling and quark Yukawa couplings, we perform a tree-level matching around the compactification scale as
\begin{align}
g_W &= g_{SU(2)_W}\nn\\
\frac{1}{2}g_W^2 &= y_t^2+y_b^2 \ \ \ \ \ \ \ \ \ \ \ \ \ {\rm at} \ \mu=\frac{1}{2\pi R},
\label{comp-weakyukawa}
\end{align}
 where $g_{SU(2)_W}$ denotes the weak gauge coupling, and the second condition is based on Eq.~(\ref{y-w}).

\begin{figure}[H]
\begin{center}
\includegraphics[width=100mm]{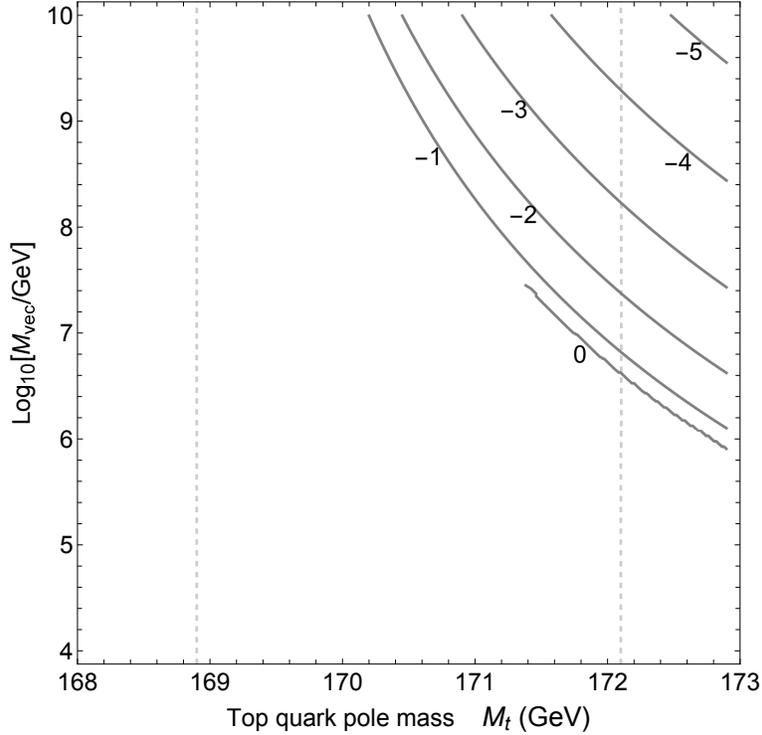}
\caption{
Contour plot of $\log_{10}\left(\mu_\lambda/\mu_{gy}\right)$ Eq.~(\ref{defratio}), which is the logarithm of the ratio of the scale where the Higgs quartic coupling vanishes and the scale where the quark Yukawa couplings and weak gauge coupling satisfy Eq.~(\ref{comp-weakyukawa}).
The vertical axis is the logarithm of the vector-like fermion mass $\log_{10}(M_{\rm vec}/{\rm GeV})$, and the horizontal axis is the top quark pole mass $M_t$.
The range of the horizontal axis corresponds to the 3$\sigma$ range of the top quark pole mass measured by the CMS Collaboration,
 and the range between the two vertical dashed lines corresponds to the 2$\sigma$ range.
On the contour of $\log_{10}\left(\mu_\lambda/\mu_{gy}\right)=0$, Eqs.~(\ref{comp-quartic}) and (\ref{comp-weakyukawa}) are simultaneously satisfied at one scale
 and the matching is successful.
}
\label{contour}
\end{center}
\end{figure}

We numerically estimate the value of the vector-like fermion mass 
 with which Eqs.~(\ref{comp-quartic}),(\ref{comp-weakyukawa}) are simultaneously satisfied at one unique scale and the matching is successful.
This scale is then identified with the compactification scale $1/(2\pi R)$.
To this end, we solve the two-loop RG equations~\cite{Machacek:1983tz,Machacek:1983fi,Machacek:1984zw} of the 4D theory 
  (comprising the SM content and the vector-like fermions)
  by varying the vector-like fermion mass.
The masses of the vector-like quarks and leptons are assumed degenerate, for simplicity.
As for the SM parameters used as inputs of the RG equations,
 since the top quark pole mass significantly affects the RG evolution of the Higgs quartic coupling,
 we vary it in the 3$\sigma$ range of the latest measurement of the CMS Collaboration, $170.5\pm0.8$~GeV~\cite{Sirunyan:2019zvx}
\footnote{
The latest measurement of the top quark pole mass of the ATLAS Collaboration is $171.1^{+1.2}_{-1.0}$~GeV~\cite{Aad:2019mkw},
 which favors large values compared to the CMS result.
}.
The strong gauge coupling is fixed at the central value of the same measurement, and 
 the other parameters are fixed at their central values in accordance with Ref.~\cite{Haba:2020ebr}.
The pole masses of the top quark, Higgs particle, and $W,Z$ particles are translated into the input values of 
 the top Yukawa coupling, the Higgs quartic coupling and the weak mixing angle for the RG equations
 by using the code~\cite{Kniehl:2016enc}, based on the
 results of \cite{Jegerlehner:2001fb,Jegerlehner:2002em,Jegerlehner:2003py,Jegerlehner:2003sp,Bezrukov:2012sa,Marquard:2015qpa,Kniehl:2015nwa}.
In Fig.~\ref{contour}, we give a contour plot of
 \bea
\log_{10}\left(\mu_\lambda/\mu_{gy}\right)
 \label{defratio}
 \eea
where
\bea
&&\lambda\left(\mu_{\lambda}\right) \ = \ 0,
\\
&&\frac{1}{2}g_{SU(2)_W}^2|_{\mu=\mu_{gy}} \ = \ y_t^2+y_b^2|_{\mu=\mu_{gy}},
\eea
  on the plane of the vector-like fermion mass $M_{\rm vec}$ versus the top quark pole mass $M_t$.
The contour of $\log_{10}\left(\mu_\lambda/\mu_{gy}\right)=0$ is
 the region where Eqs.~(\ref{comp-quartic}) and (\ref{comp-weakyukawa}) are simultaneously satisfied at one scale and the matching is successful.
Thus, this contour is the model's prediction on the vector-like fermion mass $M_{\rm vec}$ and the precise top quark pole mass.
Since the contour exists only for $M_t\gtrsim171.4$~GeV, if future measurements of the top quark pole mass exclude this range of $M_t$,
 the present model is falsified.
To confirm the model, one should test the relation between the vector-like fermion mass and the top quark pole mass given by the contour.
This is not possible in near-future collider experiments, since the vector-like fermion mass is predicted to be above $10^3$~TeV.
On the contour of $\log_{10}\left(\mu_\lambda/\mu_{gy}\right)=0$, 
 the value of $\mu_\lambda=\mu_{gy}$, which corresponds to the compactification scale $1/(2\pi R)$,
 varies from $10^{11.6}$~GeV to $10^{10.2}$~GeV from the upper-left to the lower-right.
The above range is the model's prediction on the compactification scale.

\begin{figure}[H]
\begin{center}
\includegraphics[width=100mm]{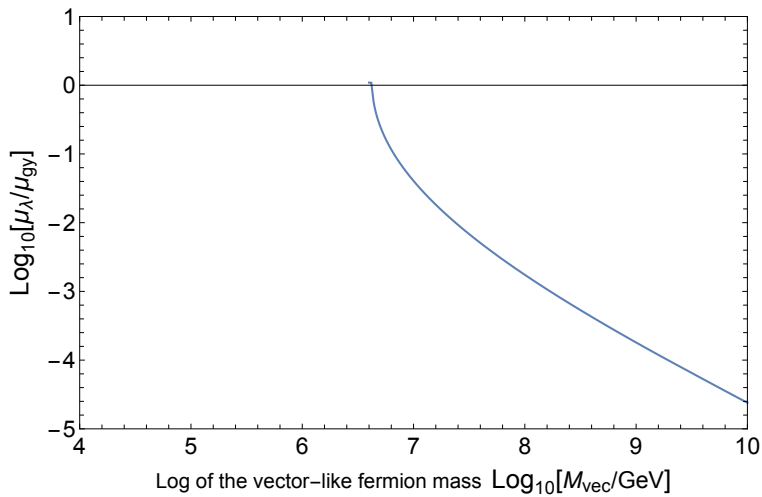}
\caption{
$\log_{10}\left(\mu_\lambda/\mu_{gy}\right)$ versus the logarithm of the vector-like fermion mass, for a fixed value of the top quark pole mass
 $M_t=172.1$~GeV.
The plot line vanishes for lower values of the vector-like fermion mass because $\mu_\lambda$ is not defined.
}
\label{fixedmt}
\end{center}
\end{figure}

\begin{figure}[H]
\begin{center}
\includegraphics[width=100mm]{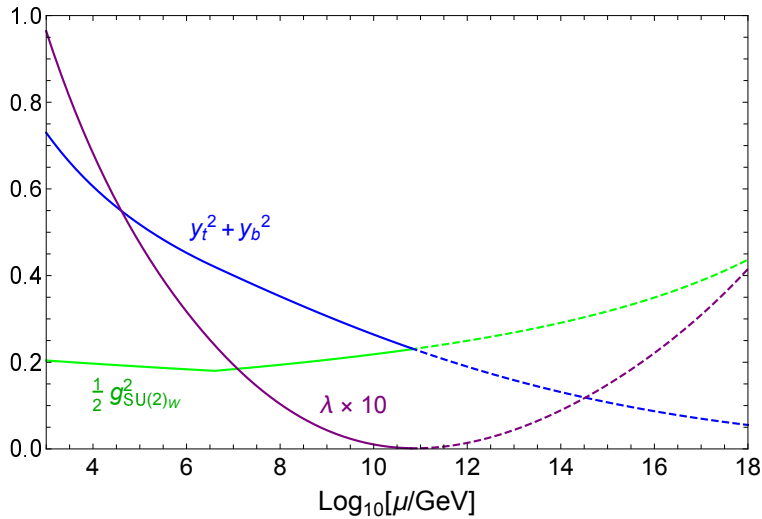}
\caption{
RG evolutions of the Higgs quartic coupling $\lambda$, the square sum of Yukawa couplings $y_t^2+y_b^2$, 
 and half the weak gauge coupling squared $(1/2)g_{SU(2)_W}^2$ for $M_t=172.1$~GeV and $\log_{10}(M_{\rm vec}/{\rm GeV})=6.6$,
  colored in purple, blue and green, respectively.
$\mu$ denotes the renormalization scale.
The solid lines are the RG evolutions in the current model, and the dashed lines are those if the 4D theory were valid at high scales.
}
\label{rgesample}
\end{center}
\end{figure}

The contour of $\log_{10}\left(\mu_\lambda/\mu_{gy}\right)=0$ is interrupted at $M_t\simeq171.4$~GeV 
 because for lower values of the top quark pole mass,
 the Higgs quartic coupling remains positive along RG evolutions
 and hence $\mu_{\lambda}$ is not defined.
There are no contours in the area under the contour of $\log_{10}\left(\mu_\lambda/\mu_{gy}\right)=0$
 because the Higgs quartic coupling remains positive there.
The reason that the Higgs quartic coupling tends to remain positive for smaller values of the vector-like fermion mass
 is that the presence of the vector-like fermions increases the weak gauge coupling along RG evolutions, which changes the beta function of the Higgs quartic coupling positive.
To visualize the situation, we show in Fig.~\ref{fixedmt} the plot of $\log_{10}\left(\mu_\lambda/\mu_{gy}\right)$ for a fixed value of the top quark pole mass
 $M_t=172.1$~GeV.
One sees that as the vector-like fermion mass decreases, $\log_{10}\left(\mu_\lambda/\mu_{gy}\right)$ increases and becomes about 0,
 and then the plot line vanishes because $\mu_\lambda$ is not defined for lower values of the vector-like fermion mass.
Interestingly, those parameter sets which yield $\mu_\lambda/\mu_{gy}=1$ are critical in the sense that 
 the Higgs quartic coupling nearly remains positive for such sets.
To illustrate this, we present in Fig.~\ref{rgesample} 
 the RG evolutions of the Higgs quartic coupling, the square sum of Yukawa couplings and the weak gauge coupling
 for a parameter set $(M_t,\,\log_{10}(M_{\rm vec}/{\rm GeV}))=(172.1$~GeV$,\, 6.6)$.
The solid purple, blue and green lines are the RG evolutions of the Higgs quartic coupling, the Yukawa coupling square sum, and half the weak gauge coupling squared, respectively. The dashed lines are their RG evolutions if the 4D theory were valid at high scales.
One confirms $\mu_\lambda/\mu_{gy}=1$ and that the Higgs quartic coupling nearly remains positive.
For such parameter sets, even a slight decrease in the vector-like fermion mass enhances the weak gauge coupling and the Higgs quartic coupling along RG evolutions, rendering the Higgs quartic coupling non-vanishing.

\section{Summary}
\label{summary}

The vanishing of the Higgs quartic coupling and the proximity of the top Yukawa and weak gauge couplings
 at high energy scales hint at 5D GHU.
Based on the above idea,
 we have proposed a model where 
 the scale of vanishing Higgs quartic coupling and 
 equality of the square sum of the top and bottom Yukawa couplings $y_t^2+y_b^2$ and half the weak gauge coupling squared $(1/2)g_{SU(2)_W}^2$,
 is interpreted as the compactification scale of 5D GHU.
The model is characterized by the embedding of the SM fermions in {\bf 3} representation of the $SU(3)_W$ gauge group.
This embedding leads to the equality of $y_t^2+y_b^2$ and $(1/2)g_{SU(2)_W}^2$
 that must be satisfied at the compactification scale.
The model structure necessitates the presence of isospin-doublet vector-like quarks and leptons.
These vector-like fermions modify the RG evolutions of the 4D theory and reconcile the scale of vanishing Higgs quartic coupling and that where  
 $y_t^2+y_b^2=(1/2)g_{SU(2)_W}^2$ holds, thereby achieving the successful matching of the 4D theory with 5D GHU.
Based on this property,
 we have calculated the RG equations of the 4D theory
 and obtained a prediction for the vector-like fermion mass, the precise top quark pole mass, and the compactification scale.
The prediction for the former two is given by the 0 contour in Fig.~\ref{contour}, 
 and that for the compactification scale varies from $10^{11.6}$~GeV to $10^{10.2}$~GeV as one moves from the upper-left to the lower-right on the contour.
About experimental testability of the model, if future measurements of the top quark pole mass exclude the region above about $171.4$~GeV,
 the model is ruled out.
The model can be confirmed by testing the relation between the vector-like fermion mass and the top quark pole mass given by the 0 contour in Fig.~\ref{contour}, 
 but this is not possible in near-future collider experiments, since the vector-like fermion mass is predicted to be above $10^3$~TeV.
\\

\section*{Acknowledgement}
This work is partially supported by Scientific Grants by the Ministry of Education, Culture, Sports, Science and Technology of Japan
 Nos.~17K05415 and 21H000761 (N.H.) and No.~19K147101 (T.Y.),
 and by the United States Department of Energy Grant No.~DE-SC0012447 (N.O.).
\\

\section*{Appendix: Matching Condition for the Higgs Quartic Coupling}

We re-derive the one-loop matching condition for the Higgs quartic coupling in Ref.~\cite{Haba:2005kc} utilized in the main text,
\bea
\lambda\left(\frac{1}{2\pi R}\right) \ = \ 0.
\nn
\eea
\\

First, we study a general $SU(3)_W$ gauge theory in a five-dimensional spacetime compactified on
 $S^1/Z_2$ orbifold, with 5D bulk fermions in the fundamental representation, 
 and without 4D chiral fermions localized at $y=0$ or $\pi R$.
The action with the gauge-fixing term is given by
\begin{align}
S=\int{\rm d}^4x\int^{\pi R}_{-\pi R}{\rm d}y \ &-\frac{1}{2}{\rm tr}\left[W_{MN} W^{MN}\right] 
- \frac{1}{\xi}{\rm tr}\left[(\partial_\mu W^\mu - \xi \overline{D}_5W_5)^2\right]
\nn\\
&+2\, {\rm tr}\left[b(\partial_\mu D^\mu - \xi \overline{D}_5 D_5) c\right]
\nn\\
&+\frac{i}{2}\overline{\Psi}\,\Gamma^MD_M\Psi - \frac{i}{2}\left(D_M\overline{\Psi}\right)\Gamma^M\Psi,
\label{action2}
\end{align}
 where $\xi$ is the gauge-fixing parameter, $b,c$ are ghost fields, and $\Psi$ represents multiple bulk fermions
 in the {\bf 3} representation of $SU(3)_W$ gauge group.
Also, in the first line, $\overline{D}_5$ denotes a covariant derivative whose gauge field part is replaced by the VEV of $W_5$ field.
We restrict ourselves to the case when $\Psi$ is $Z_2$-even, 
 as $Z_2$-odd bulk fermions give the same contribution as $Z_2$-even ones to the one-loop effective potential.
The gauge field, ghost fields and bulk fermions transform under the $Z_2$ as
\begin{align}
W_\mu(y)&=PW_\mu(-y)P^\dagger , \ \ \ \ \ W_5(y)=-PW_5(-y)P^\dagger,
\nn\\
b(y)&=P b(-y)P^\dagger , \ \ \ \ \ c(y)=P c(-y)P^\dagger,
\nn\\
\Psi(y) &= \gamma_5P\,\Psi(-y),
\nn\\
& \ \ \ \ \  \ \ P=\begin{pmatrix} 
      -1& 0 & 0 \\
      0 & -1 & 0 \\
            0 & 0 & 1 \\
   \end{pmatrix}.
   \label{z22}
\end{align}
The $Z_2$ transformation property above breaks $SU(3)_W$ into $SU(2)_W \times U(1)_W$.
The massless mode of the $SU(3)_W/(SU(2)_W \times U(1)_W)$ component of $W_5$ is identified with a Higgs field.
Unlike Ref.~\cite{Haba:2005kc}, we do not impose any non-trivial condition for the periodic change $y \to y+2\pi R$.

We compute the one-loop effective potential for the $W_5$ VEV.
To this end, we extract the quadratic part of the action Eq.~(\ref{action2}) in the presence of the $W_5$ VEV
 parametrized as
\bea
\langle W_5\rangle \ = \ \frac{1}{2gR}\begin{pmatrix} 
      0 & 0 & 0 \\
      0 & 0 & a \\
      0 & a & 0 \\
   \end{pmatrix}.
   \label{w5vev}
\eea
The quadratic part is found to be
\begin{align}
S|_{\rm quad}=&\int{\rm d}^4x\int^{\pi R}_{-\pi R}{\rm d}y \ {\rm tr}\left[
 W_\mu\left(\eta^{\mu\nu}\square-\left(1-\frac{1}{\xi}\right)\partial^\mu\partial^\nu\right)W_\nu - W_\mu\, \ov{D}_5^2 \,W^\mu \right.
\nn\\
& \ \ \ \ \ \ \ \ \ \ \ \ \ \ \ \ \ \ \ \ \ \ \ \left.-W_5\square W_5 + \xi \, W_5\, \ov{D}_5^2\, W_5 + b\square c - \xi \, b \,\ov{D}_5^2\, c \right]
\nn\\
& \ \ \ \ \ \ \ \ \ \ \ \ \ \ \ \ \ \ \ \ \ \ \ +i\overline{\Psi}\left(\gamma^\mu \partial_\mu +i \gamma_5 \ov{D}_5\right)\Psi
\label{quad1}\\
+&2\left[
\int{\rm d}^4x \ {\rm tr}\left[
2(\partial^\mu W_\mu)W_5 + W_\mu \ov{D}_5 W^\mu - \xi \, W_5 \ov{D}_5 W_5 \right] 
- \ov{\Psi}\gamma_5\Psi
\right]^{y=\pi R}_{y=0}
\label{quad2}
\end{align}
 where $\ov{D}_5$ is a covariant derivative whose gauge field part is given by $\langle W_5\rangle$ of Eq.~(\ref{w5vev}).
From the $Z_2$ transformation property and the periodicity, the fields obey the following boundary conditions at $y=0,\pi R$:
\begin{align}
W_\mu(y)&=PW_\mu(y)P^\dagger, \ \ \ \ \ W_5(y)=-PW_5(y)P^\dagger,
\nn\\
\ov{D}_5W_\mu(y)&=-P\ov{D}_5W_\mu(y)P^\dagger, \ \ \ \ \ W_5(y)=P\ov{D}_5W_5(y)P^\dagger,
\nn\\
b(y)&=Pb(y)P^\dagger, \ \ \ \ \ c(y)=Pc(y)P^\dagger,
\nn\\
\ov{D}_5b(y)&=-P\ov{D}_5b(y)P^\dagger, \ \ \ \ \ \ov{D}_5c(y)=-P\ov{D}_5c(y)P^\dagger,
\nn\\
\Psi(y) &= \gamma_5P\,\Psi(y),
\nn\\
\ov{D}_5\Psi(y) &= -\gamma_5\ov{D}_5P\,\Psi(y)
\ \ \ \ \ \ \ \ \ \ \ \ \ \ \ \ \ \ \ \ \ \ \ \ \ \ \ \ \ \ \ \ \ \ \ \ \ \ \ \ \ \ {\rm at} \ y=0,\,\pi R.
\label{bcappendix}
\end{align}
Note that given the above boundary conditions, the boundary action Eq.~(\ref{quad2}) vanishes, and the ghost term is Hermite.
We calculate the eigenvalues of the quadratic operators in the bulk Eq.~(\ref{quad1})
 under the condition that the eigenfunctions fulfill the boundary conditions Eq.~(\ref{bcappendix}).
The eigenvalue equations are given by
\begin{align}
\left(\eta^{\mu\nu}\square-\left(1-\frac{1}{\xi}\right)\partial^\mu\partial^\nu\right)W_\nu - \ov{D}_5^2 \,W^\mu 
 &= \lambda_{W_\mu} \, W^\mu,
 \nn\\
-\square W_5 + \xi \, \ov{D}_5^2\, W_5 &= \lambda_{W_5} \, W_5,
\nn\\
\square c - \xi \, \ov{D}_5^2\, c &= \lambda_c \, c,
\nn\\
\square b - \xi \, \ov{D}_5^2\, b &= \lambda_b \, b,
\nn\\
\left(i\gamma^\mu \partial_\mu - \gamma_5 \ov{D}_5\right)^2\Psi &= \lambda_\Psi \, \Psi.
\end{align}
The solutions to the above equations, {\it before} the boundary conditions are imposed, are in the form,
\begin{align}
W_\mu &= 
   e^{\pm i\,p\,x} \ \Omega(y)\left\{
   C^{TL}_\mu \cos(m\,y) + S^{TL}_\mu \sin(m\,y)
   \right\}\Omega^\dagger(y)
\nn\\
& \ \ \ \ \ \ \ \ \ \ \  + e^{\pm i\,p\,x} \ \Omega(y)\left\{
   C^S_\mu \cos(m\,y) + S^S_\mu \sin(m\,y)
   \right\}\Omega^\dagger(y)
\nn\\
W_5 &= 
   e^{\pm i\,p\,x} \ \Omega(y)\left\{
   C_5 \cos(m\,y) + S_5 \sin(m\,y)
   \right\}\Omega^\dagger(y)
   \nn\\
c &= e^{\pm i\,p\,x} \ \Omega(y)\left\{
   C_c \cos(m\,y) + S_c \sin(m\,y)
   \right\}\Omega^\dagger(y)
      \nn\\
b &= e^{\pm i\,p\,x} \ \Omega(y)\left\{
   C_b \cos(m\,y) + S_b \sin(m\,y)
   \right\}\Omega^\dagger(y)
   \nn\\
\Psi &= e^{\pm i\,p\,x} \ \Omega(y)\left\{
   C^L \cos(m\,y) + S^L \sin(m\,y)
   \right\}
+e^{\pm i\,p\,x} \ \Omega(y)\left\{
   C^R \cos(m\,y) + S^R \sin(m\,y)
   \right\},
   \label{solend}
\end{align}
where
\bea
\Omega(y)=\exp\left[i\,g\int_0^y\langle W_5\rangle {\rm d}y'
\right]
   =\exp\left[\frac{i}{2}\begin{pmatrix} 
      0 & 0 & 0 \\
      0 & 0 & a \\
      0 & a & 0 \\
   \end{pmatrix}\frac{y}{R}\right].
\eea
Here $p$ denotes a four-momentum, and $m$ denotes the mass of Kaluza-Klein modes determined from the boundary conditions.
The zero-modes correspond to the case where $m=0$.
The $C$'s and $S$'s
 are constants satisfying $p^\mu C^{TL}_\mu=p^\mu S^{TL}_\mu=0$, $C^{S}_\mu \propto S^{TL}_\mu \propto p_\mu$,
 $\gamma_5C^L=-C^L$, $\gamma_5S^L=-S^L$, $\gamma_5C^R=C^R$, $\gamma_5S^R=S^R$.
The corresponding eigenvalues are given by
\begin{align}
&\lambda_{W_\mu^{TL}}=-p^2+m^2, \ \ \ \ \ \lambda_{W_\mu^{S}}=-p^2/\xi+m^2, 
\nn\\
&\lambda_{W_5}=p^2-\xi m^2,
\nn\\
&\lambda_{c}=-p^2+\xi m^2,
\nn\\
&\lambda_{b}=-p^2+\xi m^2,
\nn\\
&\lambda_\Psi=-p^2+m^2.
\label{eigenvaluesappendix}
\end{align}

Now we impose the boundary conditions Eq.~(\ref{bcappendix}) on the solutions Eq.~(\ref{solend})
 and determine the value of $m$.
The boundary conditions are equivalent to the following constraints on the constants:
\begin{align}
&PC^{TL}_\mu P^\dagger = C^{TL}_\mu, \ \ \ \ \ \ \ \ \ \ PS^{TL}_\mu P^\dagger = -S^{TL}_\mu,
\nn\\
&P\Omega(\pi R)\left\{
   C^{TL}_\mu \cos(m\,\pi R) + S^{TL}_\mu \sin(m\,\pi R)
   \right\}\Omega^\dagger(\pi R)P^\dagger
\nn\\
&\ \ \ \ \ \ \ \ \ \ \ \ \ \ \ \ \ \ = \Omega(\pi R)\left\{
   C^{TL}_\mu \cos(m\,\pi R) + S^{TL}_\mu \sin(m\,\pi R)
   \right\}\Omega^\dagger(\pi R),
   \nn\\
&P\Omega(\pi R)\left\{
   -C^{TL}_\mu \sin(m\,\pi R) + S^{TL}_\mu \cos(m\,\pi R)
   \right\}\Omega^\dagger(\pi R)P^\dagger
\nn\\
&\ \ \ \ \ \ \ \ \ \ \ \ \ \ \ \ \ \ = -\Omega(\pi R)\left\{
   -C^{TL}_\mu \sin(m\,\pi R) + S^{TL}_\mu \cos(m\,\pi R)
   \right\}\Omega^\dagger(\pi R)
\label{constants1}\\
\nn\\
& \ \ \ \ \ \ \ \ \ {\rm (the \ same \ for \ } C^{S}_\mu,\,S^{S}_\mu,\,C_c,\,S_c,\,C_b,\,S_b, \ {\rm and \ the \ sign \ is \ flipped \ for} 
\ C_5,\,S_5),
\nn
\end{align}
 and
\begin{align}
&PC^L = -C^L, \ \ \ \ \ \ \ \ \ \ PS^L = S^L, 
\nn\\
&P\Omega(\pi R)\left\{
   C^L\cos(m\,\pi R) + S^L \sin(m\,\pi R)
   \right\}
 = -\Omega(\pi R)\left\{
   C^L\cos(m\,\pi R) + S^L \sin(m\,\pi R)
   \right\},
\nn\\
&P\Omega(\pi R)\left\{
   -C^L\sin(m\,\pi R) + S^L \cos(m\,\pi R)
   \right\}
 = \Omega(\pi R)\left\{
   -C^L\sin(m\,\pi R) + S^L \cos(m\,\pi R)
   \right\},
\nn\\
&PC^R = C^R, \ \ \ \ \ \ \ \ \ \ PS^R = -S^R, 
\nn\\
&P\Omega(\pi R)\left\{
   C^R\cos(m\,\pi R) + S^R \sin(m\,\pi R)
   \right\}
 = \Omega(\pi R)\left\{
   C^R\cos(m\,\pi R) + S^R \sin(m\,\pi R)
   \right\},
\nn\\
&P\Omega(\pi R)\left\{
   -C^R\sin(m\,\pi R) + S^R \cos(m\,\pi R)
   \right\}
 = -\Omega(\pi R)\left\{
   -C^R\sin(m\,\pi R) + S^R \cos(m\,\pi R)
   \right\}.
   \label{constants2}
\end{align}
In Eq.~(\ref{constants1}), non-zero solutions for the set $\left(C^{TL}_\mu\cos(m\,y),\,S^{TL}_\mu\sin(m\,y)\right)$ exist when
\bea
m=\frac{n}{R}, \ \ \frac{n\pm a/2}{R}, \ \ \frac{n\pm a}{R} \ \ \ \ \ \ \ \ (n=0,\pm1,\pm2,...),
\label{m1appendix}
\eea
 and when $n>0$ there are one solution for $m=n/R$, two solutions for each of $m=(n\pm \frac{a}{2})/R$,
 and one solution for each of $m=(n\pm a)/R$.
The solutions when $n<0$ are degenerate with those when $n>0$. 
When $n=0$, there are one solution for $m=0$, two common solutions for $m=\pm \frac{a}{2}/R$,
 and one common solution for $m=\pm a/R$,
 which respectively correspond to the photon, $W^\pm$ bosons, and $Z$ boson with wrong Weinberg angle.
The values of $m$ and the number of solutions for each $m$
 are the same for $(C^{S}_\mu,\,S^{S}_\mu)$, $(C_c,\,S_c)$, $(C_b,\,S_b)$, and $(C_5,\,S_5)$.
 
In Eq.~(\ref{constants2}), non-zero solutions for the set $\left(C^L\cos(m\,y),\,S^L\sin(m\,y)\right)$ exist when
\bea
m=\frac{n}{R}, \ \ \frac{n\pm a/2}{R} \ \ \ \ \ \ \ \ (n=0,\pm1,\pm2,...),
\label{m2appendix}
\eea
 and when $n>0$ there are one solution for $m=n/R$ and one solution for each of $m=(n\pm \frac{a}{2})/R$.
The solutions when $n<0$ are degenerate with those when $n>0$.
When $n=0$, there are one solution for $m=0$ and one common solution for $m=\pm\frac{a}{2}/R$,
 which respectively correspond to one Weyl fermion that does not have a Yukawa coupling with the Higgs field
\footnote{
In the model of the main text, this fermion gains a vector-like mass with a 4D fermion localized at an orbifold fixed point.
}
 and one of a pair of Weyl fermions that have a Yukawa coupling.

Non-zero solutions for the set $\left(C^R\cos(m\,y),\,S_R\sin(m\,y)\right)$ exist when
\bea
m=\frac{n}{R}, \ \ \frac{n\pm a/2}{R} \ \ \ \ \ \ \ \ (n=0,\pm1,\pm2,...),
\label{m3appendix}
\eea
 and when $n>0$ there are one solution for $m=n/R$ and one solution for each of $m=(n\pm \frac{a}{2})/R$.
The solutions when $n<0$ are degenerate with those when $n>0$.
When $n=0$, there only is one common solution for $m=\pm\frac{a}{2}/R$,
 which corresponds to one of a pair of Weyl fermions that have a Yukawa coupling.

Substituting Eqs.~(\ref{m1appendix})-(\ref{m3appendix}) into Eqs.~(\ref{eigenvaluesappendix}), we obtain the true eigenvalues.

From the eigenvalues and their duplications found above, the one-loop effective potential for the $W_5$ VEV $a$ is computed as
\begin{align}
&V_{\rm eff}(a) - V_{\rm eff}(0)
\nn\\
&=-\frac{i}{2}\int \frac{{\rm d}^4 p}{(2\pi)^4}
\sum_{n=1}^\infty\left[3 \left\{2\log\frac{-p^2+(n+\frac{a}{2})^2/R^2}{-p^2+n^2/R^2}+2\log\frac{-p^2+(n-\frac{a}{2})^2/R^2}{-p^2+n^2/R^2} \right. \right.
\nn\\
&\left. \ \ \ \ \ \ \ \ \  \ \ \ \ \ \ \ \ \  \ \ \ \ \ \ \ \ \  \ \ \ \ \ \  \ \ \ \ \ \ \ \ \  \ \ \  \ \ \ \ \ \ \ \ \ +\log\frac{-p^2+(n+a)^2/R^2}{-p^2+n^2/R^2}+\log\frac{-p^2+(n-a)^2/R^2}{-p^2+n^2/R^2} \right\}
\nn\\
&\left.\ \ \ \ \ \ \ \ \ \ \ \ \ \ \ \ \ \  \ \ \ \ \ \ \ \ \    -2N_\Psi \left\{2\log\frac{-p^2+(n+\frac{a}{2})^2/R^2}{-p^2+n^2/R^2}+2\log\frac{-p^2+(n-\frac{a}{2})^2/R^2}{-p^2+n^2/R^2}\right\}\right]
\nn\\
&\ \ \ \ \ \ \ \ \ \ \ \ \ \ \ \ \ \ \ 
+3\left\{2\log\frac{-p^2+(\frac{a}{2})^2/R^2}{-p^2}+\log\frac{-p^2+a^2/R^2}{-p^2} \right\}
-4N_\Psi \log\frac{-p^2+(\frac{a}{2})^2/R^2}{-p^2}.
\end{align}
After summation over $n$ and a Wick rotation with $p^0=ip_E^0$, one gets
\begin{align}
&V_{\rm eff}(a) - V_{\rm eff}(0)
\nn\\
&=\frac{1}{2}\int \frac{{\rm d}^4 p_E}{(2\pi)^4}
3 \left\{
2\log
\left(1+\frac{\sin^2(\pi \,\frac{a}{2})}{\sinh^2(\pi R\,p_E)}\right)
+\log\left(1+\frac{\sin^2(\pi \,a)}{\sinh^2(\pi R\,p_E)}\right)\right\}
\nn\\
&\ \ \ \ \ \ \ \ \  \ \ \ \ \ \ \ \ \  \ \ \ \ \ \ \ \ \ -4N_\Psi
\log
\left(1+\frac{\sin^2(\pi \,\frac{a}{2})}{\sinh^2(\pi R\,p_E)}\right),
\label{effpot}
\end{align}
 where $p_E^2$ denotes the Wick-rotated momentum squared, and $p_E$ denotes its square root.
Note that the above effective potential is finite.

We are interested in the Higgs quartic coupling derived from the effective potential Eq.~(\ref{effpot}).
We have a formula,
\begin{align}
&\frac{{\rm d}^4}{{\rm d}a^4}
\int \frac{{\rm d}^4 p_E}{(2\pi)^4}\log\left(1+\frac{\sin^2(\pi \,a)}{\sinh^2(\pi R\,p_E)}\right)
\ = \ \frac{3}{4\pi^2}\frac{1}{R^4}\log\left(4\sin^2(\pi\,a)\right).
\label{derivative}
\end{align}
Applying Eq.~(\ref{derivative}) to Eq.~(\ref{effpot}), we find that the quartic coupling for the physical Higgs boson $h$ is derived as
\begin{align}
&\frac{{\rm d}^4}{{\rm d}h^4}V_{\rm eff}(a) = (g R)^4\frac{{\rm d}^4}{{\rm d}a^4}V_{\rm eff}(a)
\nn\\
&\ \ \ \ =\frac{3}{8\pi^2}g^4
\left\{3\cdot 2\cdot \frac{1}{2^4}\log\left(4\sin^2(\pi\,\frac{a}{2})\right)+3\log\left(4\sin^2(\pi\,a)\right)
-4N_\Psi\cdot \frac{1}{2^4}\log\left(4\sin^2(\pi\,\frac{a}{2})\right)\right\}.
\label{quarticfinal}
\end{align}
This should be compared with the Higgs quartic coupling derived from the one-loop effective potential of
 a 4D model that contains
 $W^\pm$ bosons with mass $m=\frac{a}{2}/R$, $Z$ boson with mass $m=a/R$, $N_{\Psi}$ Dirac fermions with mass $m=\frac{a}{2}/R$, and a Higgs boson whose quartic coupling vanishes at tree level.
In the 4D model, the Higgs quartic coupling derived from the one-loop effective potential reads
\begin{align}
\frac{{\rm d}^4}{{\rm d}h^4}V_{\rm pSM\,eff}
&=\lambda(\mu) + \frac{3}{8\pi^2}\left\{
3\cdot2\left(\frac{g}{2}\right)^4 \log \frac{\frac{1}{4}g^2h^2}{\mu^2}
+3g^4 \log \frac{g^2h^2}{\mu^2} - 4N_\Psi \left(\frac{g}{2}\right)^4 \log \frac{\frac{1}{4}g^2h^2}{\mu^2} \right\},
\label{quarticpsm}
\end{align}
 where $\mu$ denotes the renormalization scale and $\lambda(\mu)$ is the Higgs quartic coupling generated from RG evolutions.
In Eq.~(\ref{quarticfinal}), when $a\ll1$ but $\log a$ is still $O(1)$ (note that $a$ need not correspond to the true Higgs VEV 246~GeV
 when deriving the matching condition for the Higgs quartic coupling),
 one can make approximations of
 $\log(4\sin^2(\pi\,a))\simeq \log(4\pi^2a^2)$ and $\log(4\sin^2(\pi\,\frac{a}{2}))\simeq \log(\pi^2a^2)$,
 while perturbation theory remains valid.
Given these approximations, 
 and noting that $h$ in Eq.~(\ref{quarticpsm}) is related to $a$ as $h=a/(gR)$,
 we obtain from the comparison of Eqs.~(\ref{quarticfinal}),(\ref{quarticpsm})
 the following one-loop matching condition for the Higgs quartic coupling:
\bea
\lambda\left(\frac{1}{2\pi R}\right) \ = \ 0.
\nn
\eea
\\

In addition to the field content of the above general model,
 the model of the main text contains extra $U(1)_V$ gauge group that is involved in the breaking of $U(1)_W\times U(1)_V\to U(1)_Y$,
 and 4D chiral fermions localized at $y=0$ that mix with 4D-decomposed components of the 5D bulk fermions along the breaking.
However, the presence of the $U(1)_V$ gauge group and the 4D chiral fermions has negligible impact on 
 the matching condition $\lambda(\frac{1}{2\pi R})=0$, as shown below.
The scale of the $U(1)_W\times U(1)_V\to U(1)_Y$ breaking is given by the VEV of a scalar $\phi$ localized at $y=0$.
When coupling constants $A_{ij},B_{ij},C_{ij},E_{ij}$ in Eq.~(\ref{action}) are $O(1)$, we have
\bea
M_{\rm vec} \sim \langle \phi\rangle.
\eea
As seen in Section~\ref{analysis}, the model predicts $M_{\rm vec}<10^{7.5}$~GeV and 
 $1/(2\pi R)>10^{10.2}$~GeV,
 and hence there is a large hierarchy between $\langle \phi\rangle$ and the compactification scale given by
 $\langle \phi\rangle /(2\pi R) \lesssim 10^{-2.7}$.
Consequently, the mixing of the $U(1)_V$ and $U(1)_W$ gauge bosons and that of the 4D chiral fermions and 4D-decomposed components of the 5D bulk fermions, induced by $\langle \phi\rangle$,
 change the scale of $\lambda=0$ only by a negligible amount of about $10^{-2.7}$ or below.
\\



\begin{thebibliography}{99}
\bibitem{Buttazzo:2013uya}
D.~Buttazzo, G.~Degrassi, P.~P.~Giardino, G.~F.~Giudice, F.~Sala, A.~Salvio and A.~Strumia,
``Investigating the near-criticality of the Higgs boson,''
JHEP \textbf{12}, 089 (2013)
[arXiv:1307.3536 [hep-ph]].

\bibitem{Haba:2005kc}
N.~Haba, S.~Matsumoto, N.~Okada and T.~Yamashita,
``Effective theoretical approach of Gauge-Higgs unification model and its phenomenological applications,''
JHEP \textbf{02}, 073 (2006)
[arXiv:hep-ph/0511046 [hep-ph]].


\bibitem{Haba:2008dr}
N.~Haba, S.~Matsumoto, N.~Okada and T.~Yamashita,
``Effective Potential of Higgs Field in Warped Gauge-Higgs Unification,''
Prog. Theor. Phys. \textbf{120}, 77-98 (2008)
[arXiv:0802.3431 [hep-ph]].



\bibitem{Carson:2015ova}
J.~Carson and N.~Okada,
``125 GeV Higgs boson mass from 5D gauge-Higgs unification,''
PTEP \textbf{2018}, no.3, 033B03 (2018)
[arXiv:1510.03092 [hep-ph]].

\bibitem{Haba:2016xji}
N.~Haba, N.~Okada and T.~Yamada,
``Proton decay prediction from a gauge-Higgs unification scenario in five dimensions,''
Phys. Rev. D \textbf{94}, no.7, 071701(R) (2016)
[arXiv:1608.04065 [hep-ph]].






\bibitem{Manton:1979kb}
N.~S.~Manton,
``A New Six-Dimensional Approach to the Weinberg-Salam Model,''
Nucl. Phys. B \textbf{158}, 141-153 (1979)
\bibitem{Fairlie:1979at}
D.~B.~Fairlie,
``Higgs' Fields and the Determination of the Weinberg Angle,''
Phys. Lett. B \textbf{82}, 97-100 (1979)
\bibitem{Fairlie:1979zy}
D.~B.~Fairlie,
``Two Consistent Calculations of the Weinberg Angle,''
J. Phys. G \textbf{5}, L55 (1979)
\bibitem{Hosotani:1983xw}
Y.~Hosotani,
``Dynamical Mass Generation by Compact Extra Dimensions,''
Phys. Lett. B \textbf{126}, 309-313 (1983)
\bibitem{Hosotani:1983vn}
Y.~Hosotani,
``Dynamical Gauge Symmetry Breaking as the Casimir Effect,''
Phys. Lett. B \textbf{129}, 193-197 (1983)
\bibitem{Hosotani:1988bm}
Y.~Hosotani,
``Dynamics of Nonintegrable Phases and Gauge Symmetry Breaking,''
Annals Phys. \textbf{190}, 233 (1989)


\bibitem{Antoniadis:1990ew}
I.~Antoniadis,
``A Possible new dimension at a few TeV,''
Phys. Lett. B \textbf{246}, 377-384 (1990)
\bibitem{Antoniadis:1993jp}
I.~Antoniadis and K.~Benakli,
``Limits on extra dimensions in orbifold compactifications of superstrings,''
Phys. Lett. B \textbf{326}, 69-78 (1994)
[arXiv:hep-th/9310151 [hep-th]].


\bibitem{Abdalgabar:2017cjw}
A.~Abdalgabar, M.~O.~Khojali, A.~S.~Cornell, G.~Cacciapaglia and A.~Deandrea,
``Unification of gauge and Yukawa couplings,''
Phys. Lett. B \textbf{776}, 231-235 (2018)
[arXiv:1706.02313 [hep-ph]].



\bibitem{Machacek:1983tz}
M.~E.~Machacek and M.~T.~Vaughn,
``Two Loop Renormalization Group Equations in a General Quantum Field Theory. 1. Wave Function Renormalization,''
Nucl. Phys. B \textbf{222}, 83-103 (1983)
\bibitem{Machacek:1983fi}
M.~E.~Machacek and M.~T.~Vaughn,
``Two Loop Renormalization Group Equations in a General Quantum Field Theory. 2. Yukawa Couplings,''
Nucl. Phys. B \textbf{236}, 221-232 (1984)
\bibitem{Machacek:1984zw}
M.~E.~Machacek and M.~T.~Vaughn,
``Two Loop Renormalization Group Equations in a General Quantum Field Theory. 3. Scalar Quartic Couplings,''
Nucl. Phys. B \textbf{249}, 70-92 (1985)



\bibitem{Sirunyan:2019zvx}
A.~M.~Sirunyan \textit{et al.} [CMS],
``Measurement of $\mathrm{t\bar t}$ normalised multi-differential cross sections in pp collisions at $\sqrt s=13$ TeV, and simultaneous determination of the strong coupling strength, top quark pole mass, and parton distribution functions,''
Eur. Phys. J. C \textbf{80}, no.7, 658 (2020)
[arXiv:1904.05237 [hep-ex]].



\bibitem{Aad:2019mkw}
G.~Aad \textit{et al.} [ATLAS],
``Measurement of the top-quark mass in $t\bar{t}+1$-jet events collected with the ATLAS detector in $pp$ collisions at $\sqrt{s}=8$ TeV,''
JHEP \textbf{11}, 150 (2019)
[arXiv:1905.02302 [hep-ex]].


\bibitem{Haba:2020ebr}
N.~Haba, Y.~Mimura and T.~Yamada,
``Renormalizable $SO (10)$ grand unified theory with suppressed dimension-5 proton decays,''
PTEP \textbf{2021}, no.2, 023B01 (2021)
[arXiv:2008.05362 [hep-ph]].
 
 
 
 
\bibitem{Kniehl:2016enc}
B.~A.~Kniehl, A.~F.~Pikelner and O.~L.~Veretin,
``mr: a C++ library for the matching and running of the Standard Model parameters,''
Comput. Phys. Commun. \textbf{206}, 84-96 (2016)
[arXiv:1601.08143 [hep-ph]].
 
 
\bibitem{Jegerlehner:2001fb}
F.~Jegerlehner, M.~Y.~Kalmykov and O.~Veretin,
``MS versus pole masses of gauge bosons: Electroweak bosonic two loop corrections,''
Nucl. Phys. B \textbf{641}, 285-326 (2002)
[arXiv:hep-ph/0105304 [hep-ph]].
\bibitem{Jegerlehner:2002em}
F.~Jegerlehner, M.~Y.~Kalmykov and O.~Veretin,
``MS-bar versus pole masses of gauge bosons. 2. Two loop electroweak fermion corrections,''
Nucl. Phys. B \textbf{658}, 49-112 (2003)
[arXiv:hep-ph/0212319 [hep-ph]].
 
 
\bibitem{Jegerlehner:2003py}
F.~Jegerlehner and M.~Y.~Kalmykov,
``O(alpha alpha(s)) correction to the pole mass of the t quark within the standard model,''
Nucl. Phys. B \textbf{676}, 365-389 (2004)
[arXiv:hep-ph/0308216 [hep-ph]].
\bibitem{Jegerlehner:2003sp}
F.~Jegerlehner and M.~Y.~Kalmykov,
``O(alpha alpha(s)) relation between pole- and MS-bar mass of the t quark,''
Acta Phys. Polon. B \textbf{34}, 5335-5344 (2003)
[arXiv:hep-ph/0310361 [hep-ph]].
  
\bibitem{Bezrukov:2012sa}
F.~Bezrukov, M.~Y.~Kalmykov, B.~A.~Kniehl and M.~Shaposhnikov,
``Higgs Boson Mass and New Physics,''
JHEP \textbf{10}, 140 (2012)
[arXiv:1205.2893 [hep-ph]].
 
\bibitem{Marquard:2015qpa}
P.~Marquard, A.~V.~Smirnov, V.~A.~Smirnov and M.~Steinhauser,
``Quark Mass Relations to Four-Loop Order in Perturbative QCD,''
Phys. Rev. Lett. \textbf{114}, no.14, 142002 (2015)
[arXiv:1502.01030 [hep-ph]].
 
  
\bibitem{Kniehl:2015nwa}
B.~A.~Kniehl, A.~F.~Pikelner and O.~L.~Veretin,
``Two-loop electroweak threshold corrections in the Standard Model,''
Nucl. Phys. B \textbf{896}, 19-51 (2015)
[arXiv:1503.02138 [hep-ph]].
 
 


 
 
\end{thebibliography}
\end{document}